\documentclass[prl,twocolumn,showpacs,superscriptaddress,nobibnotes,nofootinbib,altaffilletter]{revtex4}

\usepackage{graphicx}
\usepackage{dcolumn}
\usepackage{bm}
\usepackage{xspace}

\begin{document}

\newcommand{\mevcsq}{$\textrm{MeV/}c^{2}$\xspace}
\newcommand{\gevcsq}{$\textrm{GeV/}c^{2}$\xspace}

\newcommand{\ks}{$K_{S}$\xspace}

\newcommand{\mkspi}{$K_{S} \pi^{+}$\xspace}
\newcommand{\mksk}{$K_{S} K^{+}$\xspace}
\newcommand{\mkpipi}{$K^{-} \pi^{+} \pi^{+}$\xspace}

\newcommand{\kspi}{$D^{+} \rightarrow 
             K_{S} \pi^{+}$\xspace}
\newcommand{\kspim}{$D^{-} 
            \rightarrow K_{S}\pi^{-}$\xspace}

\newcommand{\kpipi}{$D^{+} \rightarrow 
            K^{-}\pi^{+}\pi^{+}$\xspace}
\newcommand{\kpipim}{$D^{-} \rightarrow 
            K^{+}\pi^{-}\pi^{-}$\xspace}

\newcommand{\dpipipi}{$D^{+} \rightarrow
            \pi^{-}\pi^{+}\pi^{+}$\xspace}

\newcommand{\dstar}{$D^{*+} \rightarrow 
            D^{0}(\rightarrow K^{-}\pi^{+})\pi^{+}$\xspace}

\newcommand{\ksk}{$D^{+} \rightarrow 
            K_{S}K^{+}$\xspace}
\newcommand{\kskm}{$D^{-} \rightarrow 
            K_{S}K^{-}$\xspace}

\newcommand{\kslnu}{$D^{+} \rightarrow
            K_{S}l^{+}\nu_{l}$\xspace}

\newcommand{\kstark}{$D_{s}^{+} \rightarrow
            K^{*+}\bar{K}^{0}$\xspace}

\newcommand{\kstarbk}{$D_{s}^{+} \rightarrow
           \bar{K}^{*0}K^{+}$\xspace}

\newcommand{\akspi}{$A_{CP}
             (K_{S}\pi^+)$\xspace}
\newcommand{\aksk}{$~A_{CP}
           (K_{S}K^+)$\xspace}

\newcommand{\gkspi}{$~\Gamma(D^{+} \rightarrow 
            \bar{K}^{0}\pi^{+})$~}
\newcommand{\gkpipi}{$~\Gamma(D^{+} \rightarrow 
            K^{-}\pi^{+}\pi^{+})$~}
\newcommand{\gksk}{$~\Gamma(D^{+} \rightarrow 
            \bar{K}^{0}K^{+})$~}

\newcommand{\gresa}{$\frac{\Gamma(D^{+} \rightarrow \bar{K}^{0}\pi^{+})}
                          {\Gamma(D^{+} \rightarrow K^{-}\pi^{+}\pi^{+})}$}

\newcommand{\gresb}{$\frac{\Gamma(D^{+} \rightarrow \bar{K}^{0}K^{+})}
                          {\Gamma(D^{+} \rightarrow K^{-}\pi^{+}\pi^{+})}$}

\newcommand{\gresc}{$\frac{\Gamma(D^{+} \rightarrow \bar{K}^{0}K^{+})}
                          {\Gamma(D^{+} \rightarrow \bar{K}^{0}\pi^{+})}$}

\newcommand{\kspibr}{$(30.60 \pm 0.46 \pm 0.32)\%$}
\newcommand{\kskbr}{$( 6.04 \pm 0.35 \pm 0.30)\%$}
\newcommand{\kskbra}{$(19.96 \pm 1.19 \pm 0.96)\%$}
\newcommand{\kspiacp}{$(-1.6 \pm 1.5 \pm 0.9)\%$}
\newcommand{\kskacp}{$(+6.9 \pm 6.0 \pm 1.5)\%$}
\newcommand{\kskacpa}{$(+7.1 \pm 6.1 \pm 1.2)\%$}

\newcommand{\pdgbra}{$(32.0 \pm 4.0)\%$}
\newcommand{\pdgbrb}{$( 7.7 \pm 2.2)\%$}
\newcommand{\pdgbrc}{$(26.3 \pm 3.5)\%$}

\preprint{CU-HEP/2001-1}
\title{\boldmath Search for CP Violation in the decays \kspi and \ksk
}

\affiliation{University of California, Davis, CA 95616}
\affiliation{Centro Brasileiro de Pesquisas F\'isicas, Rio de Janeiro, RJ, Brasil}
\affiliation{CINVESTAV, 07000 M\'exico City, DF, Mexico}
\affiliation{University of Colorado, Boulder, CO 80309}
\affiliation{Fermi National Accelerator Laboratory, Batavia, IL 60510}
\affiliation{Laboratori Nazionali di Frascati dell'INFN, Frascati, Italy I-00044}
\affiliation{University of Illinois, Urbana-Champaign, IL 61801}
\affiliation{Indiana University, Bloomington, IN 47405}
\affiliation{Korea University, Seoul, Korea 136-701}
\affiliation{INFN and University of Milano, Milano, Italy}
\affiliation{University of North Carolina, Asheville, NC 28804}
\affiliation{Dipartimento di Fisica Nucleare e Teorica and INFN, Pavia, Italy}
\affiliation{University of Puerto Rico, Mayaguez, PR 00681}
\affiliation{University of South Carolina, Columbia, SC 29208}
\affiliation{University of Tennessee, Knoxville, TN 37996}
\affiliation{Vanderbilt University, Nashville, TN 37235}
\affiliation{University of Wisconsin, Madison, WI 53706}
\author{J.~M.~Link}
\affiliation{University of California, Davis, CA 95616}
\author{M.~Reyes}
\affiliation{University of California, Davis, CA 95616}
\author{P.~M.~Yager}
\affiliation{University of California, Davis, CA 95616}
\author{J.~C.~Anjos}
\affiliation{Centro Brasileiro de Pesquisas F\'isicas, Rio de Janeiro, RJ, Brasil}
\author{I.~Bediaga}
\affiliation{Centro Brasileiro de Pesquisas F\'isicas, Rio de Janeiro, RJ, Brasil}
\author{C.~G\"obel}
\affiliation{Centro Brasileiro de Pesquisas F\'isicas, Rio de Janeiro, RJ, Brasil}
\author{J.~Magnin}
\affiliation{Centro Brasileiro de Pesquisas F\'isicas, Rio de Janeiro, RJ, Brasil}
\author{A.~Massafferri}
\affiliation{Centro Brasileiro de Pesquisas F\'isicas, Rio de Janeiro, RJ, Brasil}
\author{J.~M.~de~Miranda}
\affiliation{Centro Brasileiro de Pesquisas F\'isicas, Rio de Janeiro, RJ, Brasil}
\author{I.~M.~Pepe}
\affiliation{Centro Brasileiro de Pesquisas F\'isicas, Rio de Janeiro, RJ, Brasil}
% ...
\collaboration{The FOCUS Collaboration}

\date{\today}
\begin{abstract}
A high-statistics sample of photo-produced charm from the FOCUS (E831)
experiment at Fermilab has been used to search for direct CP violation 
in the decays \kspi and \ksk. We have measured the following asymmetry
parameters relative to \kpipi: \akspi = \kspiacp,
\aksk~=~\kskacp~and 
\aksk~=~\kskacpa~relative to \kspi. The first errors
quoted are statistical and the second are systematic. We have also 
measured the relative branching ratios and found:
\gkspi/\gkpipi~=~\kspibr, 
\gksk/\gkpipi~=~\kskbr~
and
\gksk/\gkspi~=~\kskbra.
\end{abstract}

\pacs{11.30.Er 13.20.Fc 14.40.Lb}

\maketitle

CP violation occurs when the decay rate of a particle differs from that
of its CP conjugate \cite{BigiSanda}. In the Kobayashi-Maskawa ansatz this
arises due to the non-vanishing phase in the Cabibbo-Kobayashi-Maskawa
matrix when the decay amplitude has contributions from at least two 
quark diagrams with differing weak phases. In addition final state
interactions (FSI) must provide a strong phase shift.
In the Standard Model direct CP violation in the charm meson system
is predicted to occur at the level of $10^{-3}$ 
or below \cite{Buccella:1995nf}. The 
mechanism usually considered is the interference of the tree and penguin 
amplitudes in singly-Cabibbo suppressed (SCS) decays. In the 
decay \kspi the Cabibbo favored (CF) and doubly-Cabibbo suppressed (DCS) 
amplitudes contribute
coherently with, perhaps, a different weak 
phase.\footnote{The charge conjugate state is implied unless stated otherwise.}
In addition the isospin content of the DCS amplitude differs from that
of the CF case so we can expect a non-trivial strong phase shift. Several
authors have commented on the effect of $K^{0}$ mixing on the CP asymmetry
for this decay mode and the possibility of using it to search for new
physics \cite{Bigi:1995aw,Lipkin:1999qz}.

Differences in the non-leptonic decay amplitudes of charmed mesons are 
almost certainly due to FSI. These effects tend to be amplified in the 
charmed system making it an ideal laboratory for their 
study \cite{Rosner:1999xd}. The isospin amplitudes
and phase shifts in $D\rightarrow KK$, $D\rightarrow K \pi$ and 
$D\rightarrow \pi\pi$ decays can be extracted from
measurements of the branching fractions \cite{Bishai:1997km}. For example 
the magnitude of the I=3/2 amplitude can be obtained directly from the
$D^{+}\rightarrow \bar{K}^{0}\pi^{+}$ partial width \cite{Bauer:1987bm}. 

Previous studies of \kspi and \ksk have concentrated on measuring
relative branching ratios \cite{Anjos:1990nm,Frabetti:1995wu}.
This paper reports the first measurement of the CP asymmetry for 
these decays.

The data were collected during the 1996--1997 fixed target run at Fermilab.
Bremsstrahlung of electrons and positrons with an endpoint energy of 
approximately 300 GeV produces a photon beam. These beam photons interact in a
segmented beryllium-oxide
target and produce charmed particles. The average photon energy
for events which satisfy our trigger is $\simeq$~180 GeV\@. 
FOCUS uses an upgraded
version of the E687 spectrometer which is described in detail 
elsewhere \cite{Frabetti:1992au}.
Charged decay products are momentum analyzed by two oppositely polarized 
dipole magnets. Tracking is performed by a system of silicon vertex detectors
in the target region and by multi-wire proportional chambers downstream of
the interaction. Particle identification is performed by three threshold
\v{C}erenkov counters, two electromagnetic calorimeters, an hadronic 
calorimeter, and by a system of muon detectors.

The \kpipi decay is reconstructed using a candidate driven vertexing
algorithm. A decay vertex is formed using the reconstructed tracks
after which the momentum vector of the parent $D$ meson is intersected
with other tracks in the event to form a production vertex. The confidence
level of the secondary vertex is required to be greater than 1\%. 
The likelihood
for each charged particle to be an electron, pion, kaon or proton based
on the light yield from each threshold \v{C}erenkov counter is 
computed \cite{Link:2001xx}.
We demand that the Kaon hypothesis $W_{K}$, 
(i.e. $-2\ln(\text{kaon\_likelihood})$), be 
favored over the pion hypothesis $W_{\pi}$ by $\Delta W=W_{\pi}-W_{K} \ge 1$.
We also make a pion consistency cut by finding the alternative minimum 
hypothesis $W_{\text{min}}$ and requiring $W_{\text{min}}-W_{\pi}>-2$
for both pions. We
eliminate contamination due to \dstar by asking that neither $K\pi$
invariant mass combination lies within 25 \mevcsq of the nominal $D^{0}$
mass.

The techniques used for \ks reconstruction are described 
elsewhere \cite{Link:2001yy}. 
Because 90\% of \ks decays occur after the \ks has passed through
the silicon strip detector we are unable to employ the same vertexing
algorithm used to reconstruct the \kpipi decay. 
Instead we use the momentum information from
the \ks decay and the silicon track of the charged daughter
to form a candidate $D$ vector. This vector is intersected with 
candidate production vertices which are formed from two other silicon tracks.
\begin{figure}
\centering
\includegraphics[width=8.3cm]{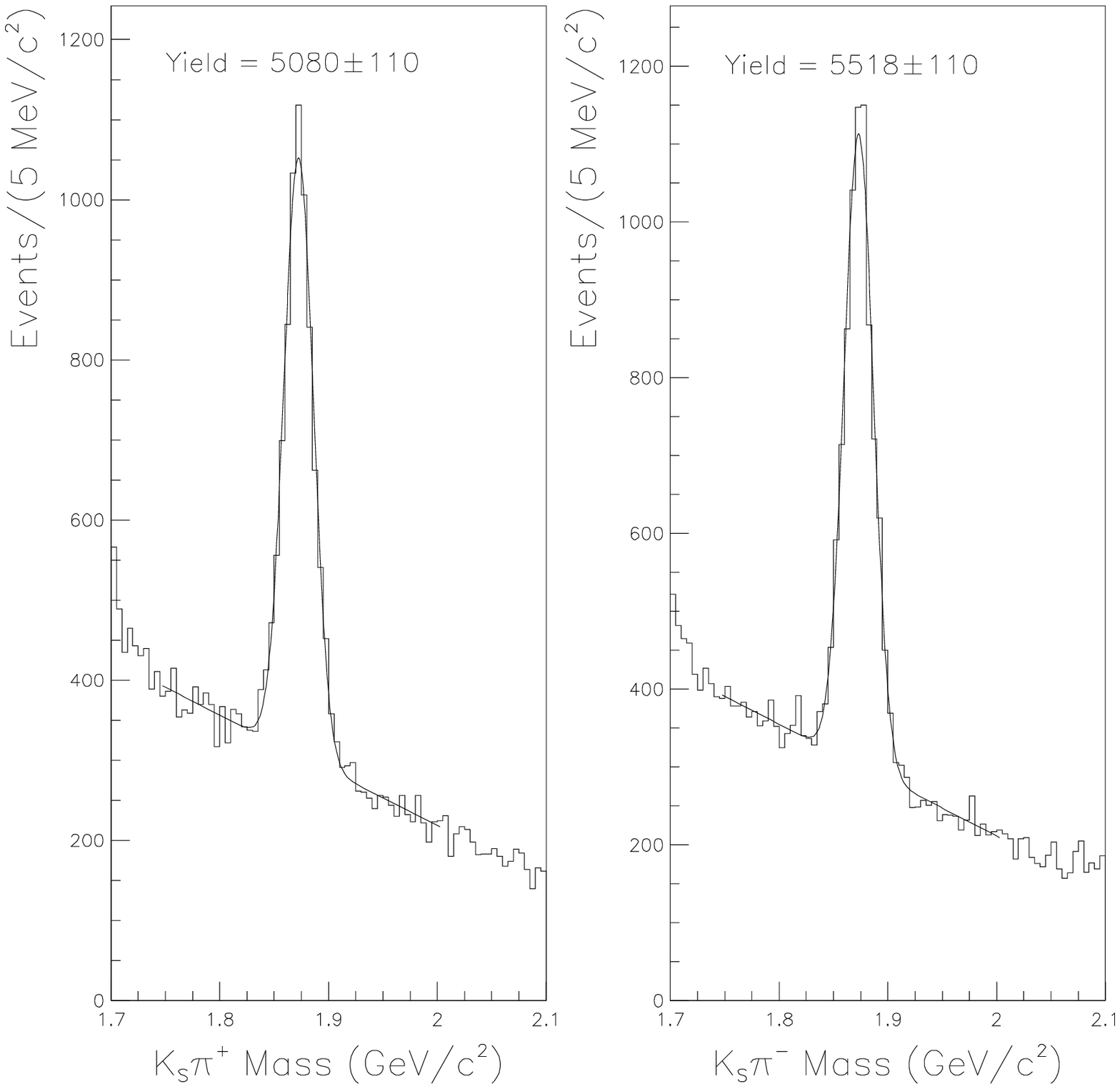}
\caption{Invariant mass plots for \kspi~and \\ \kspim}
\label{fig:kspi}
\end{figure}
As a final check we force the $D$ vector to originate at our production 
vertex and calculate the confidence level that it verticizes with the
charged daughter. This confidence level must be greater than 2\%.
We require that the momentum of the charged daughter be greater than 10
GeV/$c$, that the confidence level for it to be a muon be less than 1\%, and
that it traverse the entire length of the spectrometer. For the decay \kspi
we demand that $W_{\text{min}}-W_{\pi}>-6$ and $W_{\pi}-W_{K} < 0$, 
for \ksk we
ask that the kaon hypothesis be favored over both the proton and pion
hypotheses by requiring $W_{p}-W_{K} > 0$ and $W_{\pi}-W_{K} > 3$. We remove
electron contamination by ensuring that the charged $D$ daughter 
links to only one silicon microstrip track. Electron pairs 
usually have a very small opening angle in the silicon and chamber tracks
tend to link to both tracks. Checks for electron contamination of
the \ks sample using the electromagnetic calorimeters showed no significant
effect.
We use only \ks candidates which have a normalized mass\footnote{The 
normalized mass is the difference between the measured and the nominal mass 
divided by the error on the measured mass.} within three 
standard deviations
of the nominal value. Additionally, to reduce backgrounds in the \ksk mode,
we do not use the category of \ks decays which occur downstream of the 
silicon where both
\ks daughters lie outside the acceptance of the downstream magnet. We make 
the same cut on the \kspi normalization signal.
When the \ks decays in the silicon detector we demand that all three
tracks be inconsistent with originating at the same vertex. This 
eliminates backgrounds from decays such as \dpipipi.

\begin{figure}
\centering
\includegraphics[width=8.3cm]{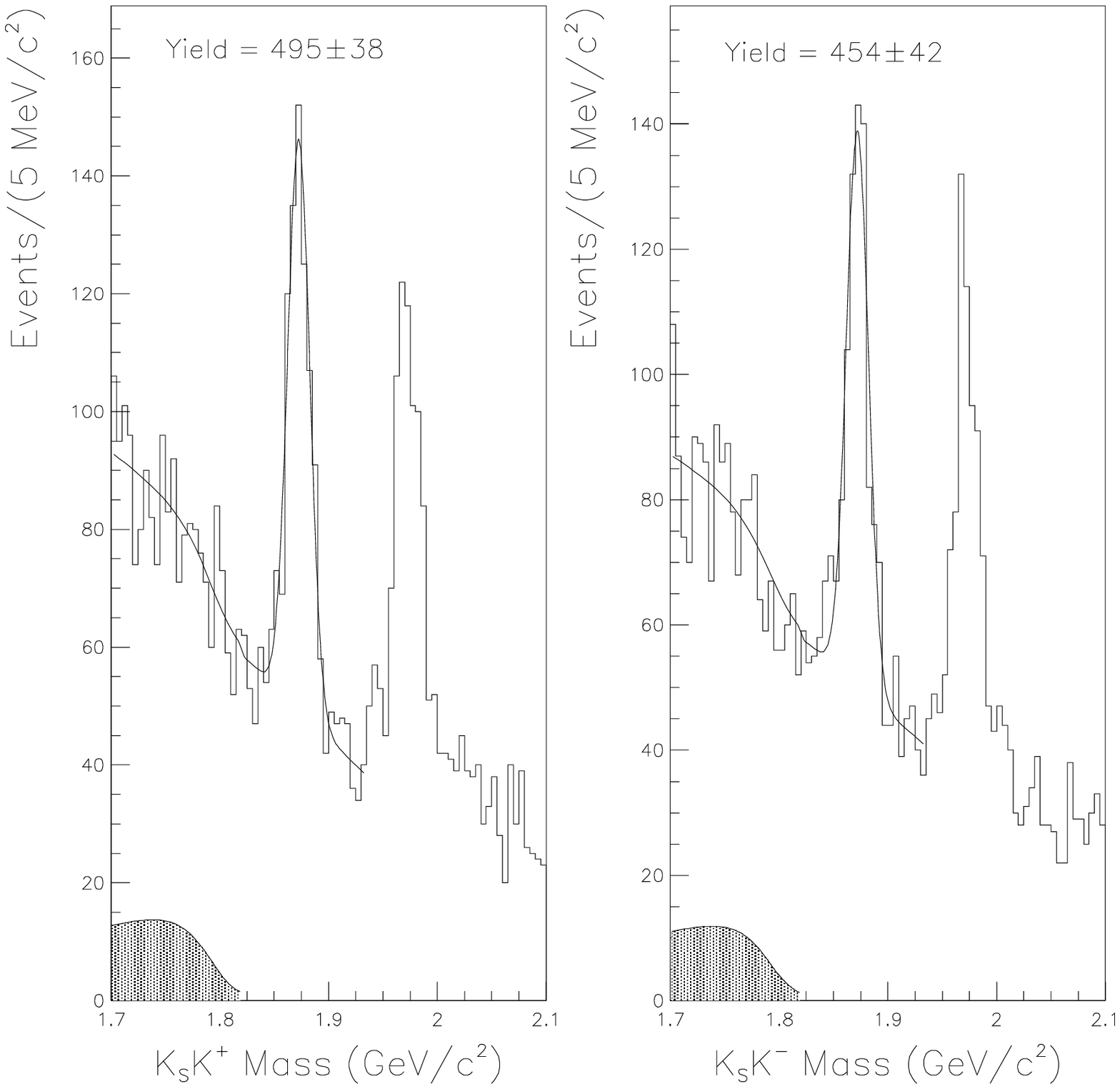}
\caption{Invariant mass plot for \ksk~and \\ \kskm. The shaded area is the
smoothed background shape from \kstark and \kstarbk.}
\label{fig:ksk}
\end{figure}

\begin{figure}
\centering
\includegraphics[width=8.3cm]{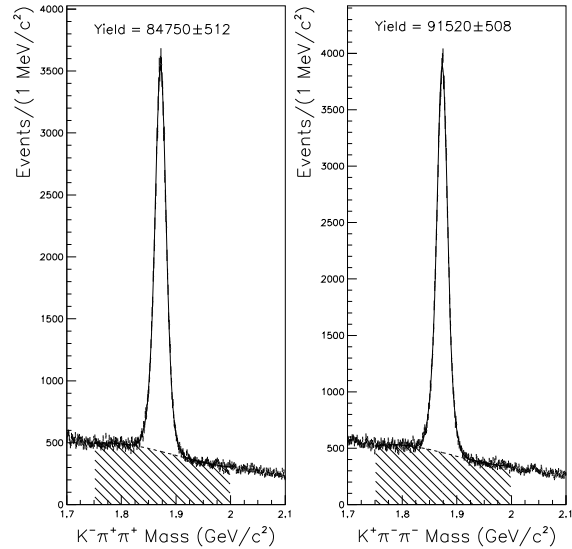}
\caption{Invariant mass plot for \kpipi~and \\ \kpipim. The shaded area is
the third-degree \\ polynomial background in the fit region.}
\label{fig:k2pi}
\end{figure}

For all modes we require that the production
vertex have a confidence level greater than 1\%, that the maximum 
confidence level for a candidate-$D$ daughter track to form a vertex with
tracks from the primary vertex be less than 20\%, that the significance of
separation of the production and decay vertices be greater than 7.5 and that
both vertices lie upstream of the first trigger counter. The momentum
of the $D$ must be greater than 40 GeV/$c$. In Figures \ref{fig:kspi},
\ref{fig:ksk} and \ref{fig:k2pi}
we show the invariant mass distributions for the decays \mkspi,
\mksk and \mkpipi respectively. \\
We construct the CP asymmetry, $A_{CP}$ as the difference in the 
yields, (corrected for efficiency and acceptance), of the decay in 
question divided by the sum. We must also
account for differences in production between the $D^{+}$ and $D^{-}$. To
do this we ratio the corrected yields to those of a Cabibbo favored decay
which is assumed to be CP conserving. We measure:
$$A_{CP} = \frac{\eta(D^{+}) - \eta(D^{-})}{\eta(D^{+}) + \eta(D^{-})}$$
where for example,
$$\eta(D^{+}) = \frac{N(D^{+} \rightarrow K_{S} \pi^{+})}
                     {N(D^{+} \rightarrow K^{-} \pi^{+} \pi^{+}) } $$
is the ratio of the corrected yields for each 
decay which is equivalent to the relative branching ratio. 

To account for non-Gaussian tails in the \kpipi
signals we find it necessary to fit these distributions using two
Gaussians and a third-degree polynomial.

The \mkspi distribution is fit using a Gaussian and a 
linear polynomial. The non-linear background shape below 
1.75 \gevcsq in the \mkspi plot is primarily due to \kslnu and 
is not included in the fit.

We fit the \mksk signal using a combination of a Gaussian, linear
polynomial, and a background shape derived from Monte Carlo. This shape
is a smoothed fit to \kstark and \kstarbk decays which, due to a missing
$\pi^{0}$, are responsible
for the background shape below the $D^+$ peak. Because of the difficulty
in fitting the region between the $D^+$ and $D_{s}^{+}$ peak in the \mksk
distribution we only fit up to 1.935 \gevcsq.
To minimize systematic errors we change the \ks selection cuts on the \kspi 
normalization signal to match those used for the \ksk mode. The yield for 
the decay \kspi changes to 4487$\pm$96 events and for \kspim
becomes 4770$\pm$96 events.
We can now calculate the relative branching ratios and CP asymmetries. The
results are shown in Tables \ref{tab:br} and \ref{tab:acp}.

We studied systematic effects due to uncertainties in our Monte Carlo 
production model, reconstruction algorithm, and variations in our 
selection cuts. For the \kspi measurements we split the sample into 
eight statistically independent subsamples based on $D^{+}$ momentum,
loose and tight normalized \ks mass cuts, and the time period in which
the data were collected.
The momentum dependence of the result arises mainly due to uncertainties
in the parameters used to generate our Monte Carlo. The \kspi topology 
and reconstruction algorithm is substantially different from that of the
\kpipi and the two modes differ in how well the Monte Carlo matches to
the data. For example there is a slight difference in how well
the generated and accepted momentum distributions agree in each case. 
We use a 
technique modeled after the {\em S-factor} method used by the Particle
Data Group \cite{Groom:2000in} to evaluate the systematic error. A 
scaled variance
is calculated using the eight independent subsamples. The split sample
systematic is defined as the difference between the scaled variance and
the statistical variance when the former exceeds the latter. Due to the
smaller statistics in the \ksk decay mode we can only form
four independent subsamples. These are based on the run period in which
the data were collected and on the normalized \ks mass.

\begin{table}
\caption{Relative branching ratio results. The first error is statistical and
the second is systematic. We  account for the decay chain
$\bar{K}^{0} \rightarrow K_{S}
\rightarrow \pi^{+}\pi^{-}$ by multiplying our \ks numbers by a factor
of 2.91 assuming that
\gkspi $=~2\times\Gamma(D^{+} \rightarrow K_{S}\pi^{+})$
; we then quote these results in terms of $\bar{K}^{0}$.}
 \begin{tabular}{rrr}
\hline
\multicolumn{1}{c}{Measurement}&\multicolumn{1}{c}{Result}&\multicolumn{1}{c}{PDG Average}\cite{Groom:2000in}\\
\hline
  \gresa & \kspibr & \pdgbra \\
  \gresb & \kskbr  & \pdgbrb\footnote{This is the measurement of reference \onlinecite{Bishai:1997km} with statistical and systematic errors added in quadrature.}\\
  \gresc & \kskbra & \pdgbrc \\
\hline
 \end{tabular}
 \label{tab:br}
\end{table}

\begin{table}
\caption{CP asymmetry measurements. The first error is statistical and
the second is systematic. }
 \begin{tabular}{ll}
\hline
   \multicolumn{1}{c}{Measurement}& \multicolumn{1}{c}{Result} \\
\hline
  \akspi~w.r.t. \kpipi  & \kspiacp \\
  \aksk ~w.r.t. \kpipi  & \kskacp  \\
  \aksk ~w.r.t. \kspi   & \kskacpa \\
\hline
 \end{tabular}
 \label{tab:acp}
\end{table}

We evaluate systematic uncertainties due to the fitting procedure 
by calculating our results for
various fit conditions, such as rebinning the histograms, changing the 
background shapes and in the case of \ksk also fitting the 
$D_{s}$ peak. Since these different results are all
{\em a priori} likely we use the resulting sample variance as a systematic.
The total systematic is calculated by adding the fit-variant systematic and
the split-sample systematic in quadrature. 
For the \kspi measurements the systematic has contributions from both the
split-sample and fit-variant analyses. For the \gkspi/\gkpipi
measurement the contribution from the 
split-sample is 0.301\% and from the fit-variant 0.098\%. For the
\akspi measurement the split-sample contribution is 0.92\% and that of
the fit-variant is 0.13\%. We find no systematic contribution to the
\ksk measurements from the split-sample technique, and therefore the
fit-variant contributions are identical to the total systematic error and
are as shown in Tables \ref{tab:br} 
and \ref{tab:acp}. Due to the lower
statistics we did not split the \ksk sample by momentum. Instead we treat
the weighted average of two samples split by momentum as a fit variant.

To conclude, we have searched for evidence of direct CP violation in the
decays \kspi and \ksk and measured their branching ratios relative to
each other and to \kpipi. Our relative branching ratios are a considerable
improvement over previous measurements. The CP asymmetries have not been
previously measured for these modes and are consistent with zero.

We wish to acknowledge the assistance of the staffs of Fermi National
Accelerator Laboratory, the INFN of Italy and the physics departments
of the collaborating institutions.
This research was supported in
part by the National Science Foundation, the U.S. Department of Energy,
the Italian Istituto Nazionale di Fisica Nucleare and
Minsitero dell'Universit\`a e della Ricerca Scientifica e Tecnologica,
the Brazilian Conselho Nacional de Desenvolvimento
Cient\'{\i}fico e Tecnol\'ogico, CONACyT-M\'exico,
the Korean Ministry of Education
and the Korean Science and Engineering Foundation.
\bibliographystyle{apsrev}
\bibliography{ksx}

\end{document}